\documentclass[apjl]{emulateapj}
\usepackage{rotating}
\usepackage{threeparttable}
 \usepackage{epstopdf}
 
\begin{document}

\title{Probing the Off-State of Cluster Giant Radio Halos}
\shorttitle{Cluster Stacking with SUMSS}
\author{S. Brown\altaffilmark{1,2}, A. Emerick\altaffilmark{3}, L. Rudnick\altaffilmark{3}, and G. Brunetti\altaffilmark{4}}

\altaffiltext{1}{CSIRO Astronomy \& Space Science; corresponding author: shea.brown@csiro.au}
\altaffiltext{2}{Bolton Fellow}
\altaffiltext{3}{University of Minnesota, 116 Church 
Street SE, Minneapolis, MN  55455}
\altaffiltext{4}{INAF - Istituto di Radioastronomia, Via P. Gobetti 101, I-40129 Bologna, Italy}

\begin{abstract} We derive the best characterization to date of the properties of radio quiescent massive galaxy clusters through a statistical analysis of their average synchrotron emissivity. We stacked 105 radio images of clusters from the 843~MHz SUMSS survey, all with L$_{X}$ $>$ 10$^{44}$~erg~s$^{-1}$ and redshifts z~$<$~0.2, after removing point-source contamination and rescaling to a common physical size. Each stacked cluster individually shows no significant large-scale diffuse radio emission at current sensitivity levels. Stacking of sub-samples leads to the following results: $(i)$ clusters with $L_{X} > 3 \times 10^{44}$~erg~s$^{-1}$ show a 6$\sigma$ detection of Mpc-scale diffuse emission with a 1.4~GHz luminosity of 2.4$\pm$0.4 $\times$10$^{23}$~W~Hz$^{-1}$. This is 1.5-2 times lower than the upper limits for radio quiescent clusters from the GMRT Radio Halo Survey (Venturi et al. 2008), and is the first independent confirmation of radio halo bi-modality; $(ii)$ clusters with low X-ray concentrations have a mean  radio luminosity (2.6$\pm$0.6 $\times$10$^{23}$~W~Hz$^{-1}$) that is at least twice that of high X-ray concentration clusters, and $(iii)$ both of these detections are likely close to the low-level ``off-state'' of GRHs in most or all luminous X-ray clusters, and not the contributions from a much smaller subset of ``on-state'' GRHs following the radio/X-ray luminosity correlation. Upcoming deep radio surveys will conclusively distinguish between these two options. We briefly discuss possible origins for the ``off-state'' emission and its implications for magnetic fields in most or all luminous X-ray clusters.
\end{abstract}


\section{ INTRODUCTION} Many massive clusters of galaxies host Mpc-scale diffuse radio emission that is not associated with active galactic nuclei (see \citealt{ferr08} for a review). This synchrotron emission implies the presence of $\mu$G magnetic fields, a population of cosmic-ray electrons/positrons (CRe$^{\pm}$), and a spatially distributed mechanism to (re)accelerate the CRe$^{\pm}$. The two most likely mechanisms for the maintenance of the  CRe$^{\pm}$ distribution are turbulent re-acceleration after a major merger \citep{brun01,petr01} and secondary production through collisions of long-lived cosmic ray protons (CRp) with thermal protons \citep{denn80,blas99}.  The CRp can be accelerated at strong shocks during cluster formation, and reside in the deep potential wells for a Hubble time \citep[see, e.g.,][for a review]{blas07}. 

The GMRT Radio Halo Survey \citep{vent07,vent08}, which surveyed a complete sample of high X-ray luminosity clusters, showed a bi-modality in the GRH population which is consistent with expectations from models based on merger-driven re-acceleration  \citep{brun09}. Those clusters that host a GRH follow a correlation between the radio and X-ray luminosities, which we designate here as the ``on-state".  The remaining ``off-state" clusters show no signs of diffuse emission even in deep GMRT images. The distinguishing characteristic between on and off-state appears to be the presence or absence of major merger activity, respectively, although outliers do exist \citep{cass10,russ11}.

Despite the difficulties that pure secondary models have in explaining recent observations \citep[e.g.,][]{brun08,donn10,brow11,jelt11}, there are reasons to expect Mpc-scale diffuse emission even in the off-state clusters. Though {\it Fermi} observations have placed upper limits on the ratio of non-thermal CRp to thermal energy densities at the level of $\sim$5\%-10\% in several clusters \citep{acke10,jelt11}, secondary CRe$^{\pm}$ from these allowed levels would still produce faint diffuse emission in the presence of $\mu$G magnetic fields. Additionally, even off-state clusters in the turbulent re-acceleration model should host steep spectrum GRHs that should be detectable with upcoming low frequency radio arrays \citep[e.g.,][]{cass06,brun08,cass10b}. 

In this Letter, we provide the most stringent constraints on the off-state synchrotron emission and GRH bi-modality by stacking images of massive, radio-quiet, clusters from the Sydney University Molonglo Sky Survey \citep[SUMSS,][]{bock99}. SUMSS is a radio survey of the Southern sky at 843~MHz, $\sim45^{\prime\prime}$ resolution, $\sim$1~mJy~beam$^{-1}$ rms noise, and is very sensitive to extended low surface-brightness emission. 
We assume $H_{o}=70$, $\Omega_{\Lambda}=0.7$, $\Omega_{M}=0.3$, and define the synchrotron spectral index $\alpha$ by $I_{\nu} \sim \nu^{-\alpha}$. 

\section{DATA ANALYSIS}

\subsection{Initial Sample Selection}  We started with a sample of luminous X-ray clusters, L$_{X}$ $>$ 10$^{44}$~erg~s$^{-1}$,  so that the clusters would be sufficiently massive to have deep potential wells, reservoirs of thermal plasma, and have had time to accumulate CRp. We further restricted the clusters  to $z < 0.2$; at that redshift,  the declination-dependent SUMSS resolution, with a maximum of  90$^{\prime\prime}$, corresponded to $\sim$295~kpc, allowing at least 3 independent beams across a 1~Mpc GRH.  From the X-ray Galaxy Clusters Database \citep[BAX,][]{sada04}, there are 149 galaxy clusters with measured X-Ray luminosity L$_{X}$ $>$ 10$^{44}$~erg~s$^{-1}$, and with redshifts $z$~$<$~0.2.

\subsection{Image Processing}  
In order to uniformly probe diffuse Mpc-scale emission through stacking, we first convolved each image as a function of redshift such that each beam is $\sim$300~kpc in diameter.  We then filtered out smaller-scale emission due to compact and slightly extended radio galaxies, using the multi-scale spatial filter of \cite{rudn02} and a box size of 600~kpc.  All images were re-scaled and resampled so that each pixel is the same physical size. After this processing, a 1~Mpc GRH of a given surface-brightness would appear the same at any redshift, and would add coherently if present in more than one (or all) of the clusters.  Statistics for each image were calculated (mean, rms, and peak fluxes) within a 1 Mpc box around the center, as well as a 1-3 Mpc annular box  for background subtraction and noise estimation. 

\subsection{Quality Control} From these 149 processed images, we needed to eliminate those that could have significant contamination from residual emission from bright point sources. To quantify the possible point source contamination, we injected synthetic 10~Jy sources into each of the images prior to the processing. The resulting stack showed $\sim0.033$~Jy~beam$^{-1}$ peak emission, or 0.33\% leakage. Point source contamination in our stack is therefore negligible compared to other noise contributions.

Marginally extended sources could contaminate the filtered images at a higher level, so we eliminated their possible influence empirically. After background removal, we plotted the diffuse, filtered flux within the 1~Mpc box against the peak unfiltered flux within 1~Mpc  (not shown). Clusters with peak fluxes below 0.25~Jy~beam$^{-1}$ did not show any correlation between compact and filtered diffuse emission, and were kept for further analysis; 105 clusters met this criterion. 

Our filtering procedure can, unfortunately, also reduce the observed flux from our target diffuse emission. To assess the amount of missing flux, we therefore injected 1~Mpc, 0.5~Mpc, 0.25~Mpc, and 0.125~Mpc gaussians into each stacked image.We recover 91\%, 60\%, 18\%, and 4\% of the total injected flux, respectively. For the luminosities listed in Table 1, we assumed 1~Mpc Gaussian halos, and thus scaled all the calculated luminosities up by 1/0.91=1.1.

We also created a control sample of 105 randomly selected fields assigned to the same redshifts as the cluster sample, and processed those in an identical manner. 

Visual inspection of the cluster images did not reveal hints of large-scale diffuse emission. We compared integrated flux measurements  from  SUMSS and from the Parkes Telescope\footnote{http://www.parkes.atnf.csiro.au/} for several large sources, and found that SUMSS is sensitive out to at least 20$^{\prime}$ scales (the closest cluster in our high-$L_{X}$ sample is z=0.048, where 1~Mpc is 17.9$^{\prime}$). We further injected low signal-to-noise gaussian halos of various sizes directly into the images, and conclude that individual 1~Mpc halos with luminosities of $P_{1.4~GHz} \approx 2\times 10^{24}$~W/Hz would be visible in SUMSS over the range of redshifts in our sample (assuming $\alpha$=1.2). 

\begin{figure}
\begin{center}
\includegraphics[width=8cm]{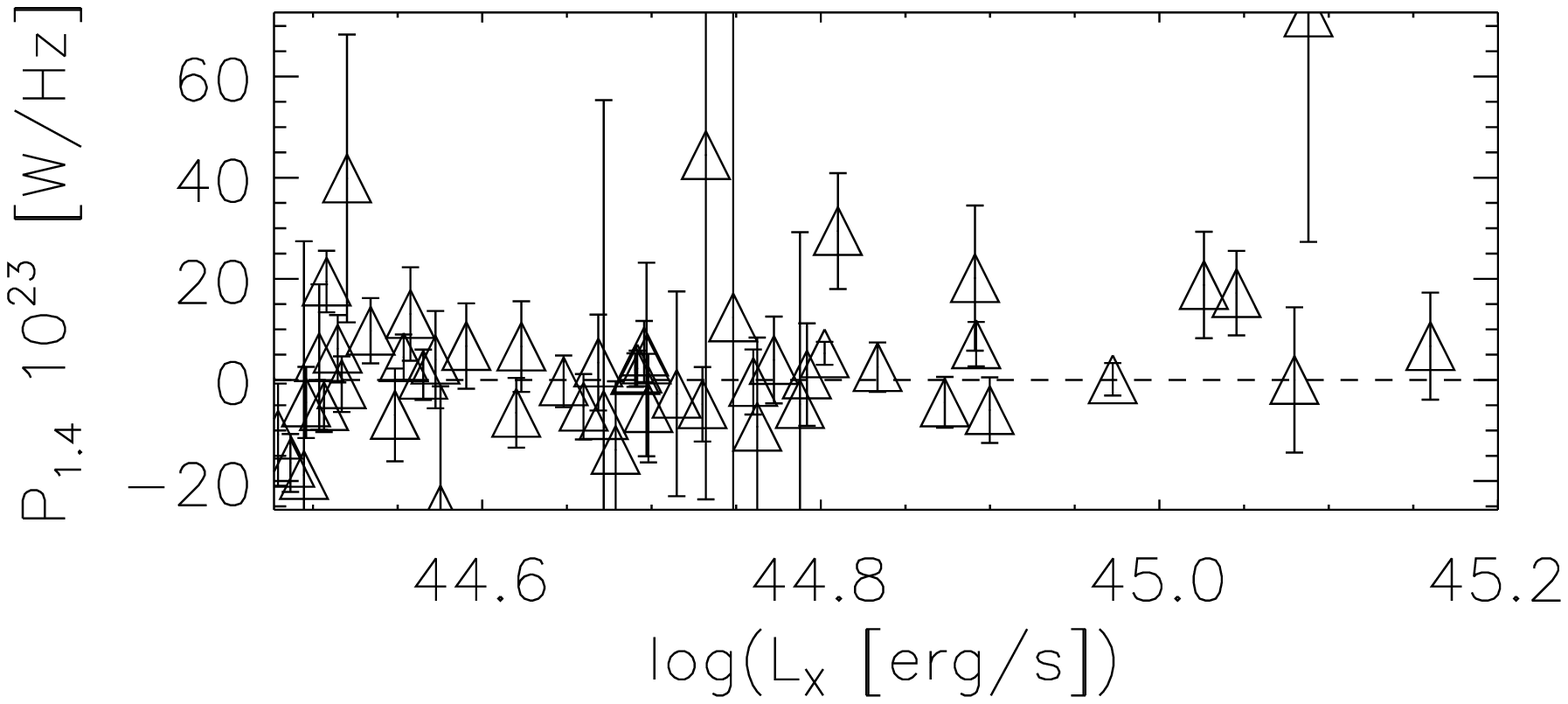}
\includegraphics[width=8cm]{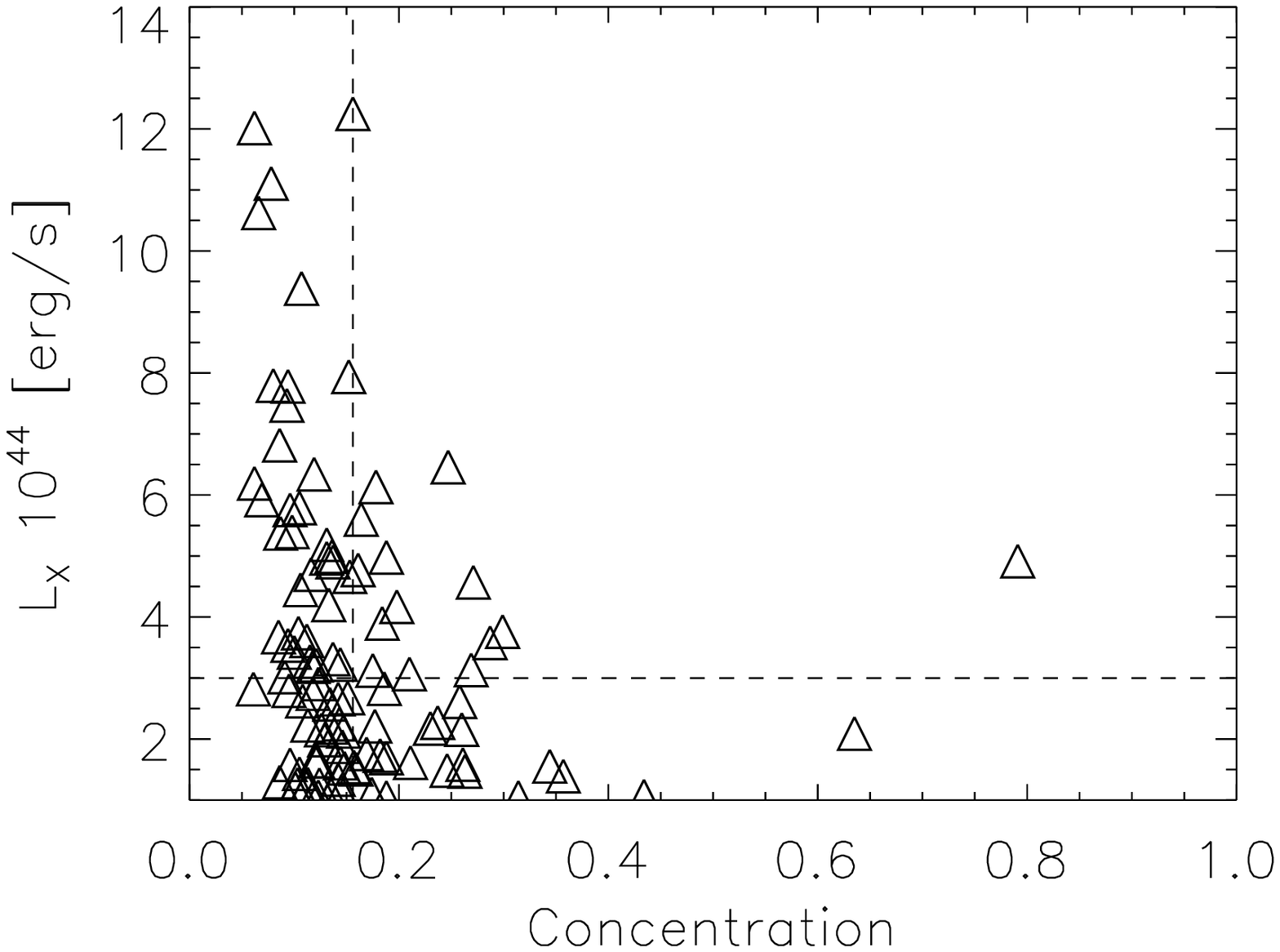}
\end{center}
\caption{\label{lrclusters} Top: Measured radio luminosities for each cluster as a function of cluster X-ray luminosity. The majority are consistent with zero (dashed line), though a few show 1-2$\sigma$ positive and negative deviations; Bottom:  A plot of cluster $L_{X}$ vs. $c$, with solid lines marking the regions of our sub-samples.}
\end{figure}

Figure \ref{lrclusters} shows the distribution of the radio luminosities (each individually a non-detection) as a function of $L_{X}$. Note that they can be either positive or negative due to the background subtraction. All radio luminosities (henceforth) are scaled to 1.4~GHz assuming an $\alpha=1.2$. 

\subsection{Stacking \& Subsamples}  
\begin{figure}
\begin{center}
\includegraphics[width=4cm]{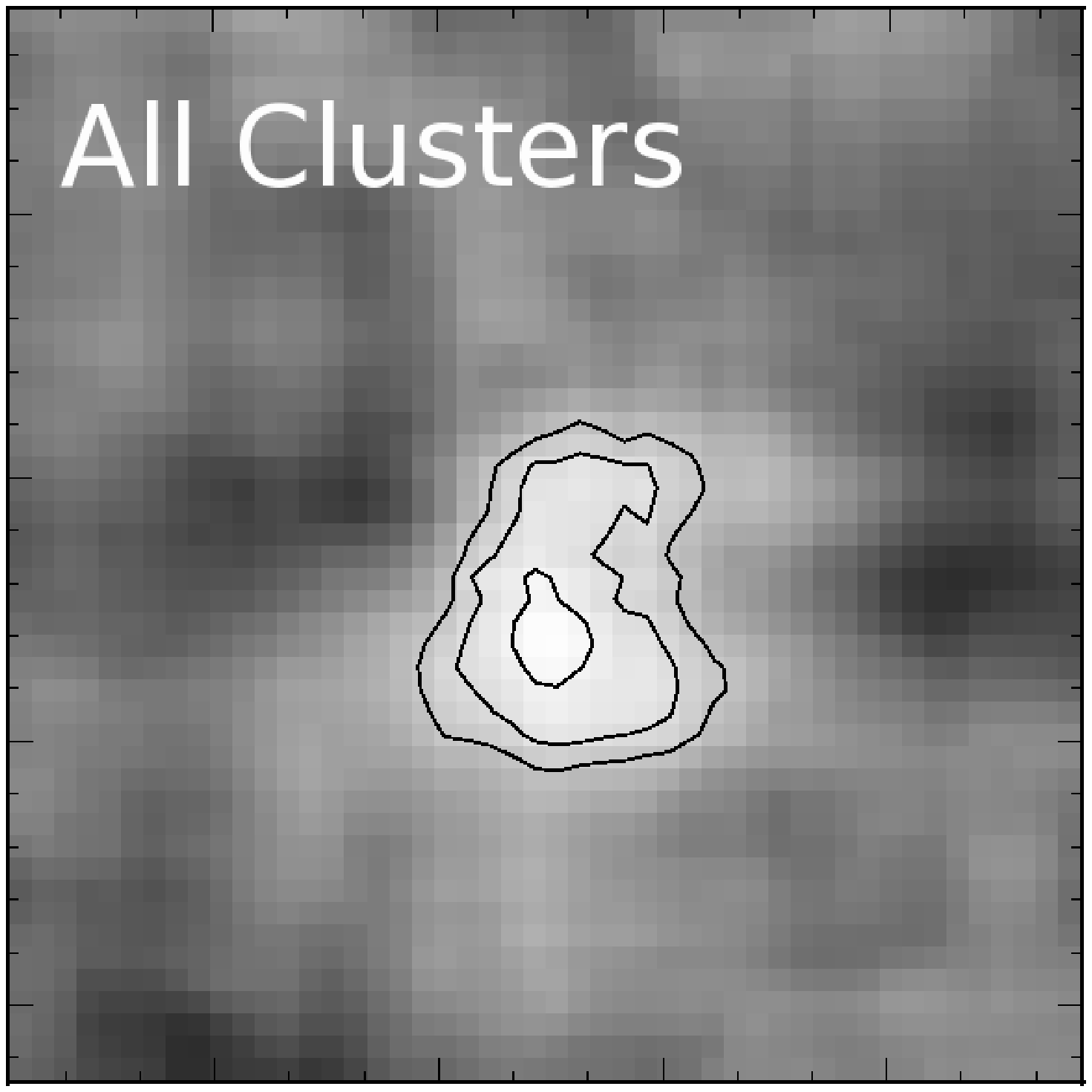}
\includegraphics[width=4cm]{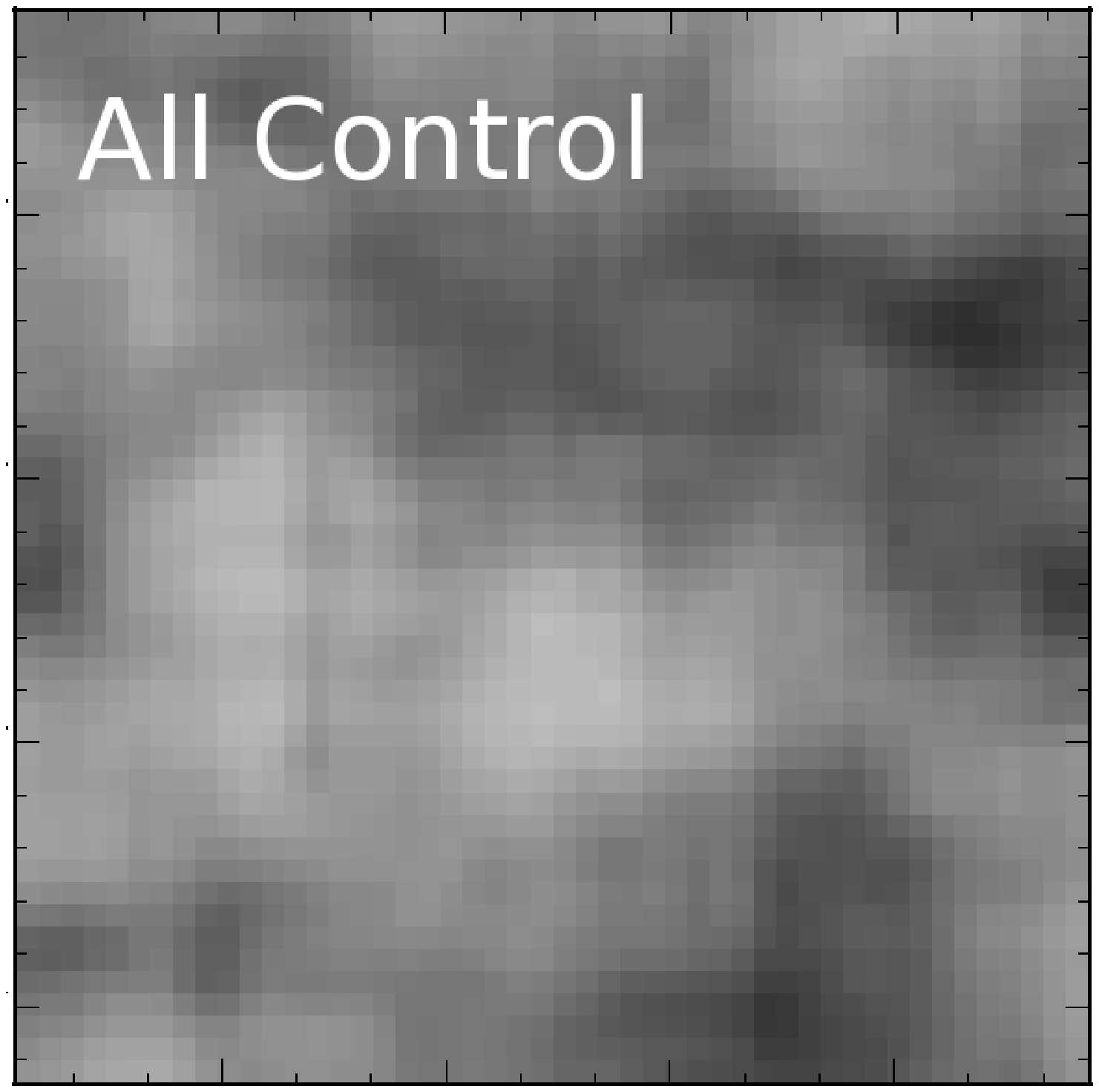}
\includegraphics[width=4cm]{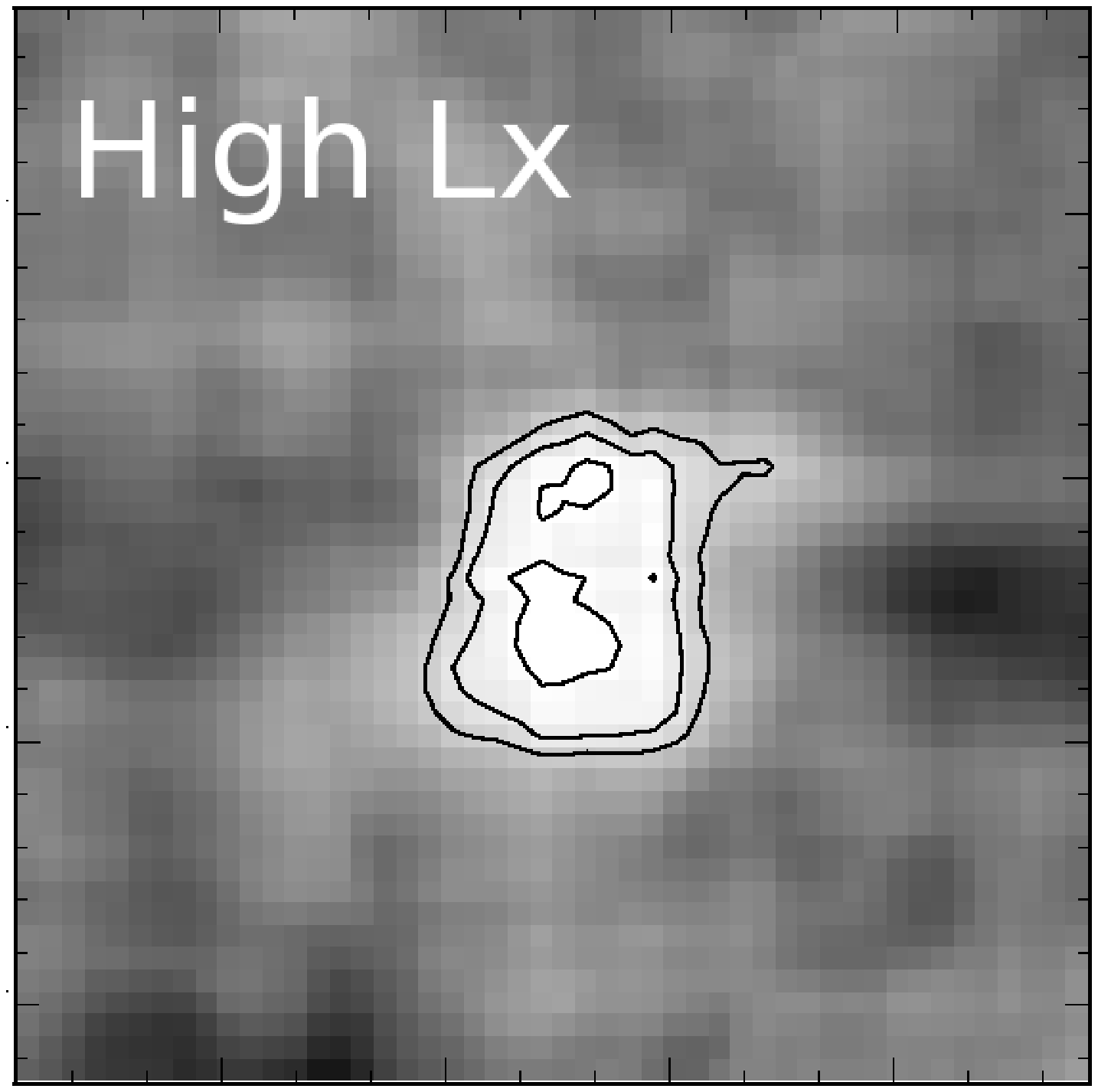}
\includegraphics[width=4cm]{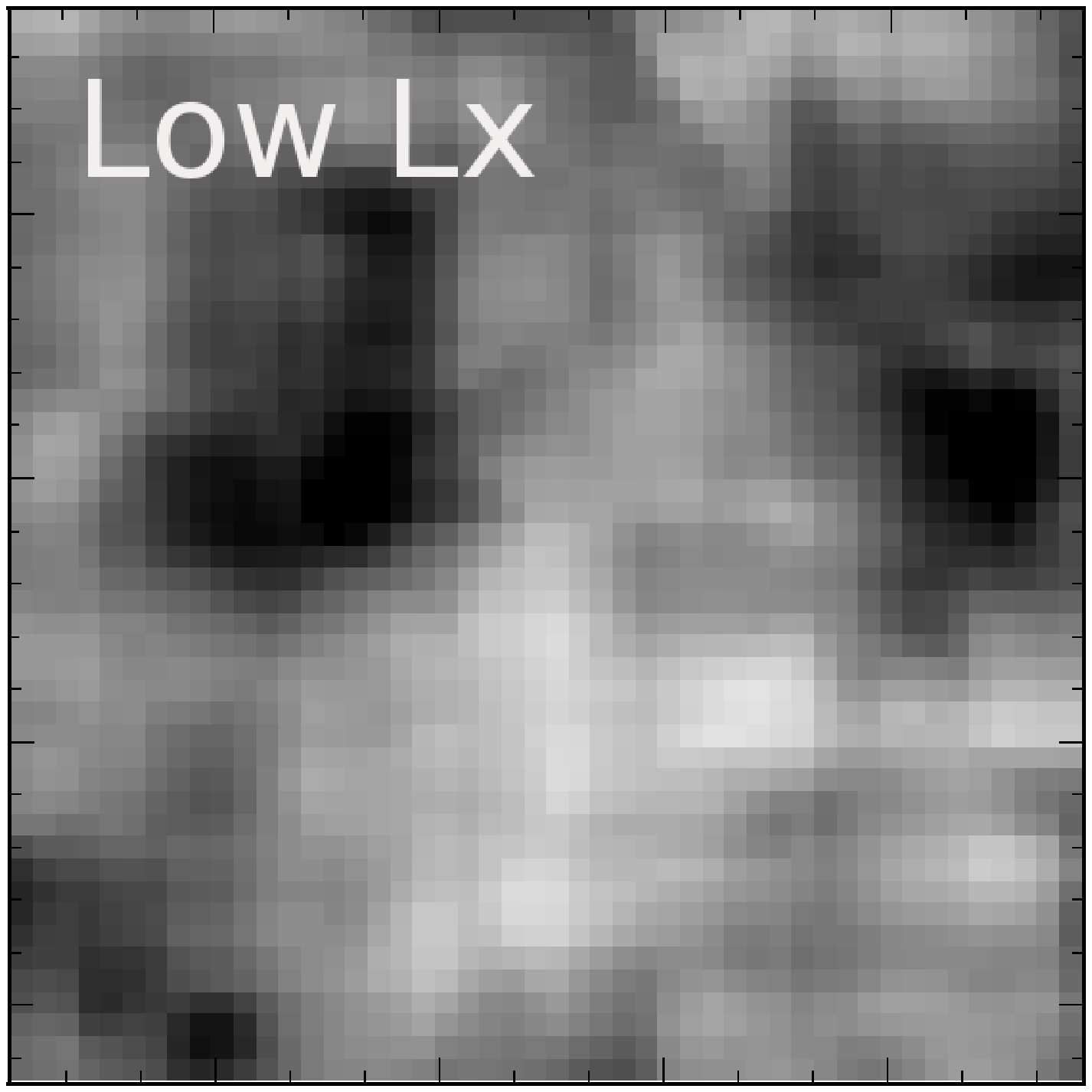}
\includegraphics[width=4cm]{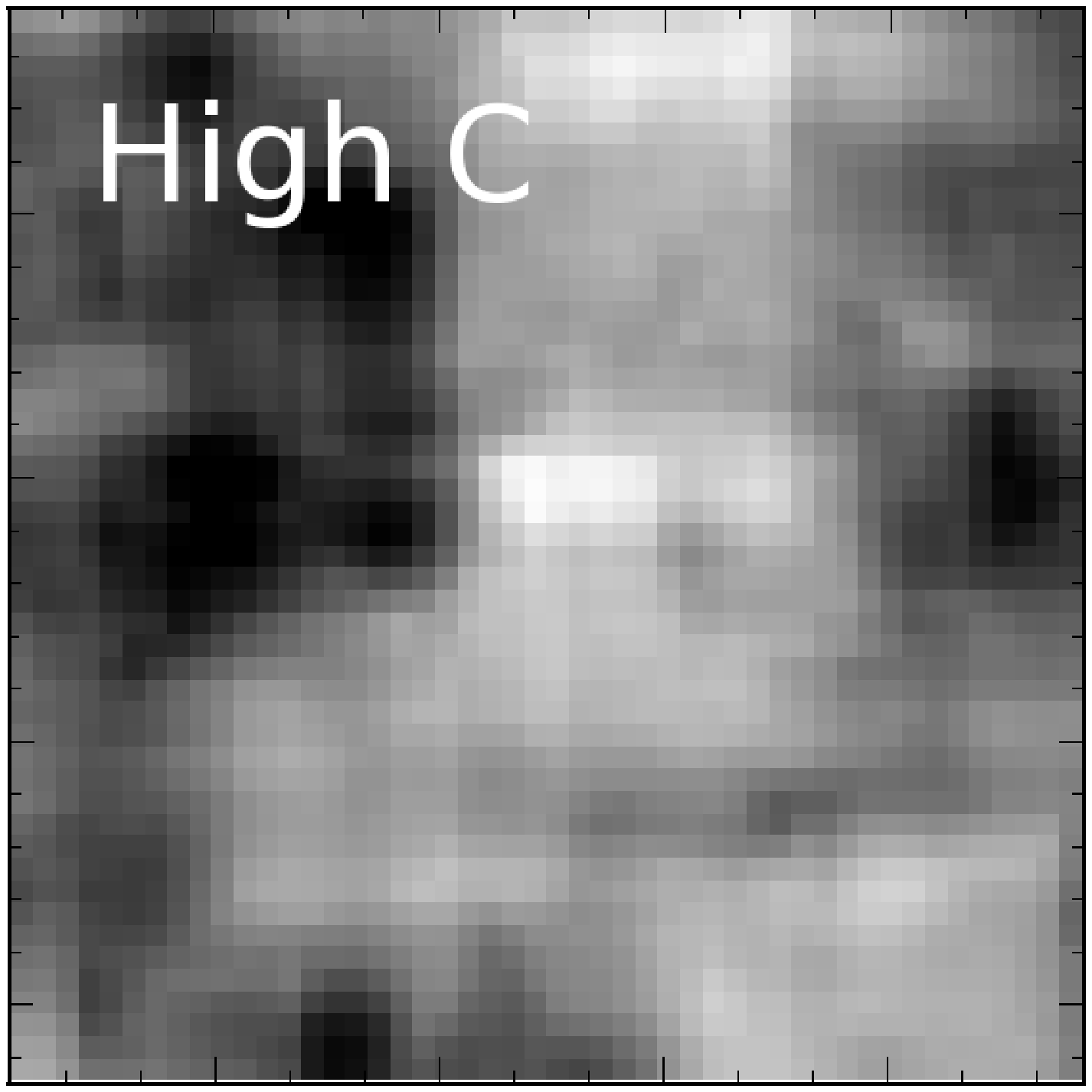}
\includegraphics[width=4cm]{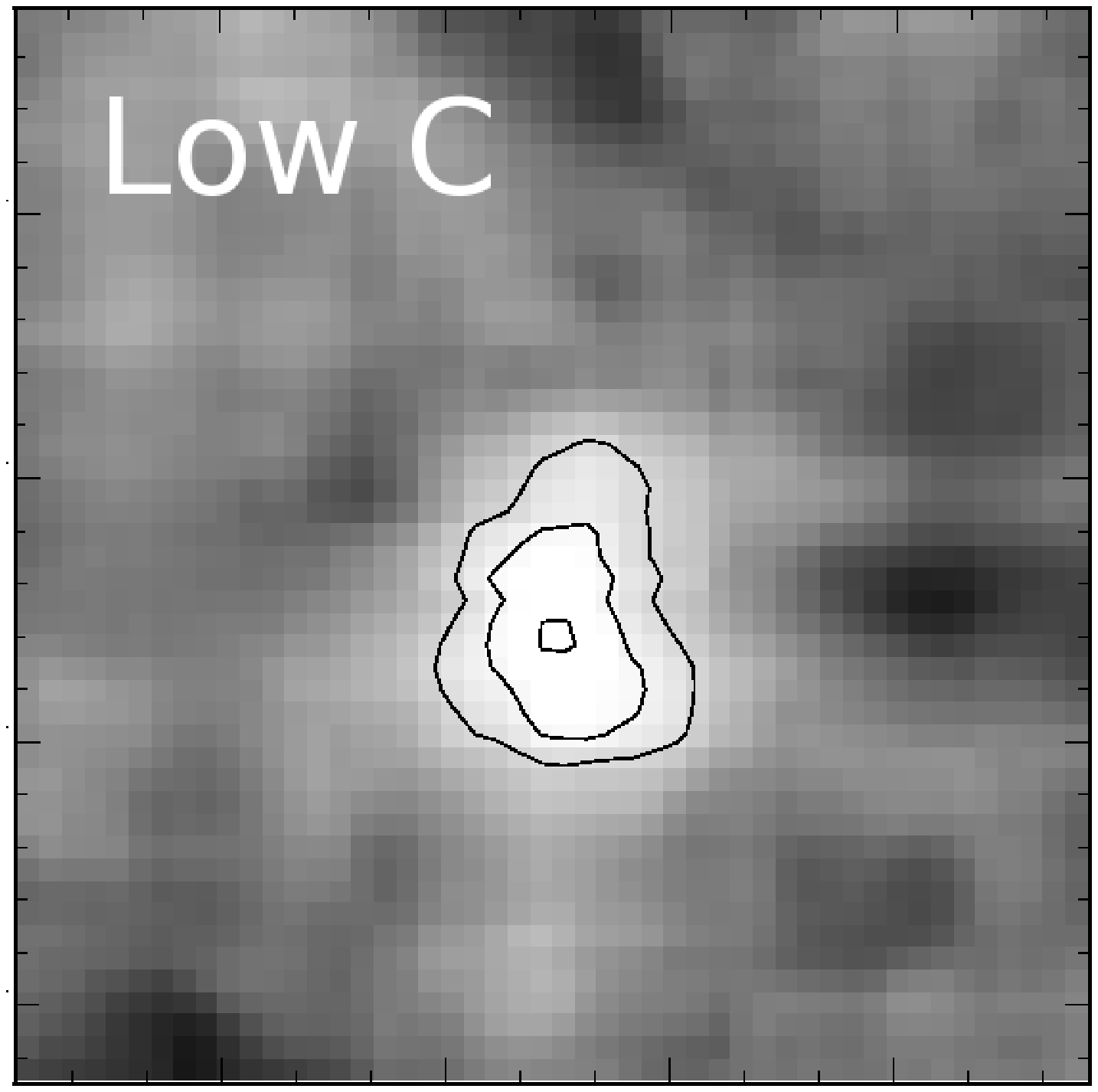}
\end{center}
\caption{\label{stacks2} For all images, greyscale runs from -1~mJy~beam$^{-1}$ (black) to +1~mJy~beam$^{-1}$ (white;
beam= 300~kpc $\times$ 300~kpc), and contours start at 3$\sigma$ and increase at 1$\sigma$ intervals. See Table 1 details of the samples.  Images are 3~Mpc on a side.}
\end{figure}

Each image had somewhat different noise properties, as well as a redshift-dependent noise due to the spatial rescaling to a common physical size.   We therefore stacked using a weighted mean, where the weight was  the inverse variance (1/$\sigma^{2}$) of each image in the background region between 1 and 3~Mpc. The results of stacking the 105 cluster and control fields are shown in Figure \ref{stacks2} and Table 1. The cluster stack shows an $\sim$4$\sigma$ detection of diffuse emission of 1~Mpc in extent, with a peak flux of 0.99$\pm$0.1~mJy~beam$^{-1}$, while the control stack is consistent with noise. 

In order to compare with theoretical expectations, we then split the 105 clusters into subsamples based on two criteria; high vs. low $L_{X}$ ($>$ or $<3 \times 10^{44}$~erg~s$^{-1}$) , and level of X-ray concentration.

To asses the concentration of X-rays we used the parameter $c$ \citep{sant08,cass10}, which is the ratio of the X-ray flux within $r<100$~kpc to that within $r<500$~kpc. Higher values of $c$ indicate more centrally dominated, and thus relaxed, clusters, while lower values of $c$  tend to indicate  the presence of sub-structure induced by cluster mergers. We calculated $c$ from the Rosat All Sky Survey \citep{voge99} images of the clusters. We ordered the individual radio luminosity values of the 104\footnote{one cluster was removed for this test due to high X-ray background resulting in $c>1$.} clusters by their $c$ parameter value and performed a Bayesian change-point analysis to determine the value of $c$ at which the distribution of luminosities changed in mean and standard deviation. The analysis found $c=0.156$ as the most likely transition value, where the ($mean, median, standard~deviation$) for clusters with $c \le 0.156$ was (5.6, 2.4, 8.2)$\times 10^{23}$~W~Hz$^{-1}$, and (2.1, 1.2, 26.0)$\times 10^{23}$~W~Hz$^{-1}$ for clusters with $c>0.156$.  We therefore use $c=0.156$ to demarcate ``merging" (low $c$)  vs. ``non-merging" clusters. We note that this is very close to the value of $c \sim 0.2$ found by \cite{cass10} as the dividing line between halo and non-halo clusters. 

\begin{figure}
\begin{center}
\includegraphics[width=8cm]{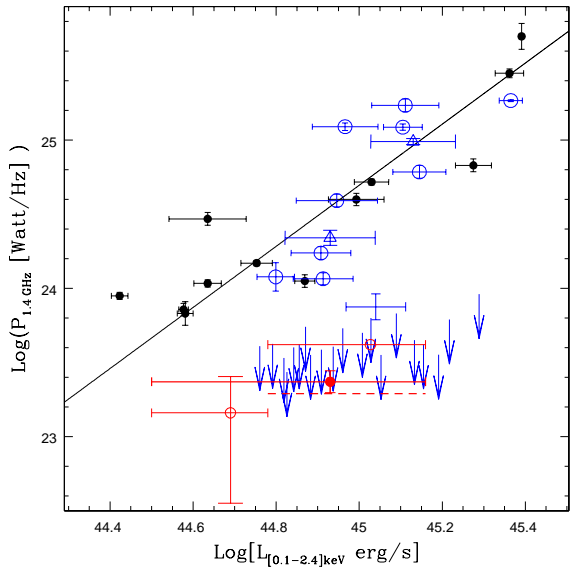}
\end{center}
\caption{\label{gmrtplot} Plot from \cite{brun09} showing the correlation between GRH's radio and X-ray luminosities. Arrows are upper limits from \cite{vent08}. The filled red circle is the stacked luminosity for the high-$L_{X}$ sample, and the dashed red line is the value with the top three contributing clusters removed.  Open red circles are results from stacking the top 13 and bottom 39 clusters in the high-$L_{X}$ sample, all assuming  1~Mpc halos. The values are placed at the weight mean $L_X$, where the weight is the predicted on-correlation value for each $L_X$ \citep{brun09}.}

\end{figure}

 Figure \ref{lrclusters} shows how the clusters are distributed in the $L_X-c$ plane. The results from subsamples characterized by either or  both of  $L_{X}$ and $c$ are shown in Figures \ref{stacks2} and Table 1. The high-$L_{X}$ and low-$c$ clusters both show a $\sim$5$\sigma$ detection of diffuse emission with 1~Mpc extent, while the low-$L_{X}$ and high-$c$ clusters are non-detections. Figure \ref{gmrtplot} shows the high-$L_{X}$ luminosity plotted along with the GMRT RHS sample in the $P-L_{X}$ plane, which is 1.5-2 times lower than the upper limits on radio quiescent clusters.

\begin{center}
\begin{threeparttable}
\centering
\caption{Stack results for sub-sampling of the 105 cluster images.}
\begin{tabular}{llcc}
\hline
$L_{X}$  [erg/s] & $c$  & N\tnote{a} & $P_{1.4~GHz}$ [10$^{23}$~W/Hz] \\
\hline
\hline
All & All & 	105 & 1.83$\pm$0.45 \\
$>3\times$10$^{44}$ & All & 52 & 2.37$\pm$0.39 \\
$<3\times$10$^{44}$ & All & 53 &  0.66$\pm$0.89\\
All & $<$0.156 & 67 & 2.62$\pm$0.66 \\
All & $>$0.156 & 37 & 1.80$\pm$1.29 \\
\hline
\end{tabular}
\label{tab:3}
\begin{tablenotes}
\item[a]{Number of clusters in the stack}
\end{tablenotes}
\end{threeparttable}
\end{center}

\section{Discussion}
 We first consider whether the origin of the diffuse emission detected at high $L_{X}$ is due to a small fraction of ordinary ``on-correlation'' radio halos that are just below the detection limit of the SUMSS survey, combined with a larger fraction of non-detections. 
To test this, we examined the distribution of radio luminosity measurements (again, each individually a non-detection) for the high X-ray luminosity sample ($L_X>3\times 10^{44}~erg/s$).  We find that 10 of the 52 high-$L_{X}$ clusters have radio errors considerably larger than the rest of the sample (see Fig. \ref{lrclusters}), so we selected the 42 lowest noise clusters and their corresponding controls and compared their luminosity distributions. Application of the Kolmogorov-Smirnov (KS) test gives a 5\% probability that the clusters are drawn from the same population as the control sample (which by definition has zero contribution of real diffuse luminosity), thus confirming the detection found in Fig. \ref{stacks2}.  The next question is whether the difference is due to a handful of luminous on-correlation clusters, or whether most/all clusters having a small positive signal far below the on-correlation.  

To address this, we first added a fixed luminosity to each control field and recalculated the probability that the cluster and control samples are drawn from the same population. The probability reaches a maximum of 91\% when we add a luminosity of 5$\times 10^{23}$~W/Hz to each control field, consistent with our measured stacked luminosity and the GMRT upper limits (Fig. \ref{gmrtplot}).  

We next performed a Monte Carlo analysis injecting an on-correlation luminosity  into a small number of the control fields, using the radio halo fractions of \cite{cass08}, and leaving the rest unchanged. After 10$^{5}$ random realisations, the mean likelihood that the observed and modified control distributions were the same was 13$\pm$5\%, with no cases $\geq$39\%.  While not statistically incontestable, we do find it most likely that  a large number of clusters have an approximate luminosity of  3-5$\times 10^{23}$~W/Hz, representing their luminosity when they are in the ``off-state". The fact that the average luminosity of an off-state cluster is an order of magnitude less than on-correlation GRHs implies a bi-modal distribution, consistent with results from the GMRT GRH Survey \citep[e.g.,][]{brun09}. 

\begin{figure}
\begin{center}
\includegraphics[width=8cm]{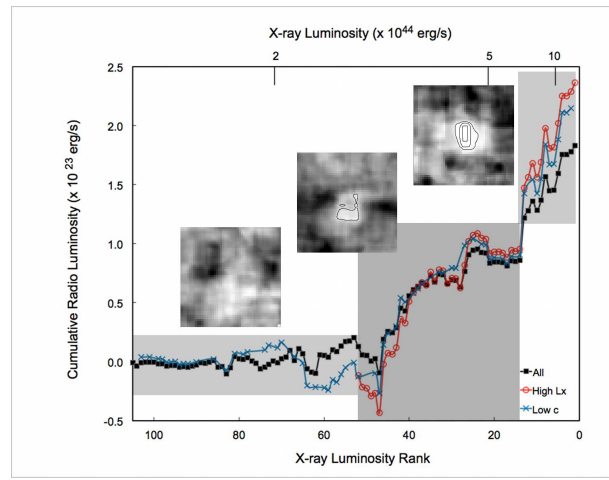}
\end{center}
\caption{\label{cumdis} Plot of the cumulative luminosity, noise-weighted as in our stacks, of the entire sample (black boxes), high-$L_{X}$ clusters (red circles), and low-$c$ clusters (blue crosses). Shaded regions indicate the sub-stacking of the top 13, middle 39, and bottom 53 clusters in the sample (see text), and the stacked images are shown as insets with the same greyscale and contours as Fig. \ref{stacks2}.}
\end{figure}

 We can further investigate how the stacked flux is distributed among the clusters by plotting the cumulative luminosity. Figure \ref{cumdis} shows the cumulative luminosity of the entire cluster sample (black squares), the high-$L_{X}$ sample (red circles), and the low-$c$ sample (blue crosses), all of which produced positive detections (Fig. \ref{stacks2}). The cumulative luminosity was calculated in ascending $L_{X}$ order, and we have applied the same weighting as in our stacking. 
There are several important features of the distribution: 1) There are several locations where the distribution jumps, indicating strong contributions from a few clusters; 2) In addition to these jumps, the overall positive trend above  $L_{X} > 3\times 10^{44}$~erg/s indicates a significant distribution of the flux among most of the clusters (as demonstrated statistically above); 3) The off-state emission is a function of $L_{X}$, even within the crude sub-sampling performed in the stacks. 

The $L_{X}$ dependence can be investigated by further splitting the high-$L_{X}$ sample into two. The top 13 clusters show a strong linear rise in cumulative luminosity, so we stacked the top 13 and the remaining 39 separately, and show the images in Fig. \ref{cumdis} and values in Fig. \ref{gmrtplot}. We see significant emission from both subsamples, again confirming that the off-state emission comes from many clusters. We note that the value obtained in the 39 cluster stack suggests that bi-modality extends down to lower $L_X$ clusters than has been probed by previous studies (see Fig. \ref{gmrtplot}). 

The statistics of GRHs observed with the GMRT suggest that, if our sample where complete, we would expect roughly 2-4 sources in the high-$L_{X}$ sample \citep{cass08}. We therefore took a more conservative approach for calculating the off-state and assumed that there could be as many as 3 GRHs in our sample. We removed the top three contributors to the cumulative distribution and recalculated the stack, obtaining $P_{1.4 GHz} = 1.95\pm0.75 \times 10^{23}$~W/Hz. We believe this value is a better estimate of the true off-state luminosity for high-$L_X$ clusters. 

The presence of low-level Mpc-scale emission in off-state clusters requires a cluster-wide CRe$^{\pm}$ acceleration mechanism and minimum  energy ICM magnetic fields of $\sim 1~\mu$G (assuming $\alpha=1.2$; \citealt{govo04}).  A similar claim of ubiquitous cluster magnetic fields is made in several studies based on rotation-measure/depolarization \citep[e.g.,][]{cari02,clar04,bona11}, although the adequate size samples of background sources needed to conduct a fair test are still missing. Possible origins for the CRe$^{\pm}$ include cluster-wide acceleration processes such as turbulent re-acceleration and CRp secondaries, or emission from diffuse radio galaxies. We consider each of these below. \\

\noindent {\sc Turbulent Re-acceleration:} There is a large body of evidence connecting the presence of GRHs with cluster mergers \citep{buot01,govo04b,cass10}. Our results (Fig \ref{stacks2}) show that disturbed clusters have stronger off-state Mpc-scale emission. However,  Fig. \ref{lrclusters} indicates that this is partially degenerate with high-$L_{X}$ for our sample.  \cite{cass10} found that on-state radio halos were not found in clusters with $c>0.2$, the radio luminosity of these systems being constrained by upper limits from GMRT radio observations at 610 MHz..  {\it Prima facie}, our results show that high $L_{X}$ and disturbed clusters preferentially show Mpc-scale diffuse emission in off-state clusters as well.  It has been suggested that radio halos with very steep spectrum are generated in connection with less energetic merger events \citep[e.g.,][]{cass06,brun08}. These sources should appear preferentially at lower radio frequencies and would be sub-luminous halos at $\sim$GHz frequencies. Thus one possibility is that our stacked radio signal from off-state clusters is from these faint systems. LOFAR is expected to detect most of these systems \citep[e.g.,][]{cass10b} and will allow for testing of this possibility. \\

\noindent {\sc CR-protons:} An exciting possibility is that we are detecting diffuse emission from secondary CRe$^{\pm}$ resulting from CRp collisions with thermal protons. {\it Fermi} limits \citep{acke10} allow the CRp energy density to be at most 5-10\% of the thermal energy density in clusters, while GMRT upper limits from the-non detection of diffuse Mpc-scale emission in off-state clusters constrain the ratio of CRp and thermal energy density to less than several percent, provided these systems are magnetized at the few $\mu$G level \citep{brun07}. The level of emission detected in these systems is close to these GMRT limits, implying a ratio of CRp to thermal energy density  $\sim$3\% (in 1 Mpc$^3$ volume), assuming $<B>\sim2~\mu$G. In a recent paper \cite{brun11} calculated models where primary CRp and their secondary products are reaccelerated by merger-driven MHD turbulence generating giant radio halos. According to these calculations the synchrotron emission from secondary particles in off-state non-turbulent systems is a factor $\sim$10 fainter than that in GRH, in rough agreement with our observations. In this scenario it is also expected that the synchrotron emission scales with the amount of turbulence in the hosting clusters, with GRH becoming gradually more powerful (and with harder spectra) in more turbulent (disturbed) systems. This may explain why we detect Mpc-scale emission preferentially in low-c clusters.  Additional possibilities for understanding the correlation with cluster dynamics might stem from differences in magnetic properties and CRp dynamics in relaxed and disturbed clusters \citep{kesh10a,kesh10,enss11}.  \\

\noindent {\sc Diffuse Radio Galaxies:} It is possible that these massive clusters host faint synchrotron emission from past/current AGN activity, though we don't see any actively driven jets on these scales in the individual images. Absent the jet driven power, the CRe$^{\pm}$ would still require some in situ re-acceleration in order to be present on these large scales. Additionally, the emission is correlated with disturbed X-ray morphology (low-$c$), while it is typically cooIing-core clusters (high-$c$) that host powerful extended radio galaxies in their core. These sources are also heavily suppressed by our filtering. 

\section{Summary} We performed a statistical analysis of the diffuse, Mpc-scale, radio emission in 105 massive clusters from the SUMSS survey. Our major findings are 1) Mpc-scale diffuse radio emission is detected in formerly radio quiescent clusters, at a level slightly below current upper limits. This emission is likely originating from clusters in the off-state; 2) there is a substantial population of clusters with $P_{1.4~GHz}$ much less than the observed $P-L_{X}$ luminosity correlation for GRHs, confirming GRH bi-modality. We further confirm bi-modality in lower $L_X$ clusters than has previously been known; 3) off-state emission is correlated with high-$L_{X}$ and disturbed cluster morphology, analogous to on-state GRHs \citep{cass10}; 4) Magnetic fields at the $\sim$1~$\mu$G level are present on Mpc-scales even in formerly radio-quiescent massive clusters.  We demonstrate the power of cluster stacking, which will allow us to probe lower mass/$L_{X}$ clusters and groups with upcoming all-sky radio continuum surveys (e.g., \citealt{norr11}).

\acknowledgements We would like to thank D. Bock for his useful comments on the SUMSS survey, and the anonymous referee for their helpful comments and suggestions. LR and AE acknowledge support from NSF grant AST0908668 to the University of Minnesota, with additional support for AE from an APS Minority Scholarship and the Minnesota REU program. GB acknowledge support from PRIN-INAF2009 and ASI-INAF I/009/10/0.


\end{document}